\documentclass[12pt]{article}

\usepackage{slashed}
\usepackage{amsmath,amssymb,graphicx} 
\usepackage{epsf}
\usepackage{pstricks}
\usepackage{cite}

\newcommand{\beq}{\begin{eqnarray}}
\newcommand{\eeq}{\end{eqnarray}}

\newcommand{\centeron}[2]{{\setbox0=\hbox{#1}\setbox1=\hbox{#2}\ifdim

\wd1>\wd0\kern.5\wd1\kern-.5\wd0\fi \copy0

\kern-.5\wd0\kern-.5\wd1\copy1\ifdim\wd0>\wd1
                                       \kern.5\wd0\kern-.5\wd1\fi}}
\newcommand{\ltap}{\>\centeron{\raise.35ex\hbox{$<$}}
                               {\lower.65ex\hbox{$\sim$}}\>}
\newcommand{\gtap}{\>\centeron{\raise.35ex\hbox{$>$}}
                               {\lower.65ex\hbox{$\sim$}}\>}

\newcommand\ZZ{\hbox{\zfont Z\kern-.4emZ}}
\font\zfont = cmss10 

\renewcommand{\theequation}{\thesection.\arabic{equation}}
\begin{document}
\begin{titlepage}

\vskip1.5cm
\begin{center}
{\huge \bf Comments on ``Gauge Fields and}
\vskip.1cm
\end{center}
\begin{center}
{\huge \bf  Unparticles"}

\vskip.1cm
\end{center}
\vskip0.2cm

\begin{center}
{\bf Jamison Galloway, Damien Martin and David Stancato} 
\end{center}
\vskip 8pt

\begin{center}
{\it Department of Physics, University of California, Davis,
CA
95616.} \\
\vspace*{0.3cm}
{\tt  galloway@physics.ucdavis.edu, martin@physics.ucdavis.edu, stancato@physics.ucdavis.edu}
\end{center}

\vglue 0.3truecm

\begin{abstract}
\vskip 3pt \noindent
The derivation of Feynman rules for unparticles carrying standard model quantum numbers is discussed.  In particular, this note demonstrates that an application of Mandelstam's approach to constructing a gauge-invariant action reproduces for unparticles the vertices one obtains through the usual minimal coupling scheme; other non-trivial requirements are satisfied as well.  This approach is compared to an alternative method that has recently been constructed by A.~L.~Licht. 

\end{abstract}

\end{titlepage}

\renewcommand{\theequation}{\thesection.\arabic{equation}}
\newcommand{\ns}{\negthinspace}
\section{Introduction}\label{sec:Introduction}
\setcounter{equation}{0}
It was recently shown by Georgi \cite{georgi1, georgi2} that non-renormalizable interactions between standard model (SM) fields and fields of a hitherto hidden conformal sector could result in missing energy in high energy collisions resembling the escape of a non-integer number of neutral particles.  In this original formulation, the fields of the conformal sector -- the `unparticles' -- do not carry SM quantum numbers, but it was speculated that providing them with charges may lead to more interesting phenomenology.  This topic has been pursued in \cite{John, Licht1, Licht2}.

The procedure for deriving the Feynman rules for gauged unparticles will be reviewed briefly below.  In this short note we hope to clarify a particular point of the derivation, viz. the utility of a particular form for the derivative of an integral operator introduced to ensure gauge invariance.  We argue that a sensible choice for the definition of this derivative is provided by the one which will reproduce the Feynman rules obtained through the usual minimal coupling prescription.  This in fact turns out to be the case in the Wilson line integral formulation if one follows the method given for regular QED  by Mandelstam \cite{Mandelstam}.  For this reason, this method is the appropriate one for unparticles from the standpoint of reproducing the familiar Feynman rules. 

Recently Licht \cite{Licht1} demonstrated that abandoning a salient postulate of Mandelstam's approach leads to the possibility of generating Feynman rules which differ from what would otherwise be expected.  Licht's approach instead involves an explicit specification of a path along which the Wilson line integral is performed.  The result of this integration is then used in the derivation of the Feynman rules.  Below we will show that the final result for the vertex function, as derived in \cite{Licht1} following this procedure, does not match the result of minimal coupling.  It is thus our aim simply to show that for unparticles, one maintains the equivalence between the method of minimal coupling and that of the use of a Wilson line supplemented by a particular specification of the directional derivative. 

In Sec.~2 we review the issue of coupling gauge fields to unparticles and the derivation of the Feynman rules.  We then show in Sec.~3 that the minimal coupling prescription is recovered for {\it particles}, i.e. in the case where the field's scaling dimension is taken to be canonical, provided the natural definition of the directional derivative is used.  Moreover, we also present an inductive proof that minimal coupling is in fact recovered for other integer values of $d$ using this approach.  We conclude in Sec.~4.

\section{Coupling Gauge Fields and Unparticles}\label{sec:Coupling Gauge Fields and Unparticles}
\setcounter{equation}{0}
In order to produce a viable model of unparticles with SM charges, one assumes an IR cutoff in the conformal sector to provide a mass gap, thus evading obvious experimental constraints.  Following Georgi, the unparticle propagators are written in spectral form, with the density function determined by simple dimensional analysis (cf. \cite{georgi1, georgi2, Cheung, John}).  Pursuing as an example the case presented in  \cite{John}, we take the propagator for scalar unparticles with scaling dimension $1\leq d<2$ to be given by the following:
\beq\label{prop}
\Delta(p,m,d) &\equiv& \int d^4x \ e^{ipx}\langle 0|T\phi (x) \phi^{\dagger}(0)|0\rangle \nonumber \\
&=& \frac{A_d}{2 \pi} \int_{m^2}^{\infty} (M^2-m^2)^{d-2} \frac{i}{p^2-M^2+i\epsilon}dM^2 \nonumber \\
&=& \frac{A_d}{2 \sin d\pi}\frac{i}{(m^2-p^2-i\epsilon)^{2-d}}.
\eeq
With proper normalization $A_d$ this reproduces Georgi's unparticle propagator as $m\rightarrow 0$ and gives the familiar scalar propagator when $d$ is taken to be canonical, i.e. $d\rightarrow 1$. 

Generically, one can view propagators such as (\ref{prop}) as coming from an effective action
\beq\label{action}
S \propto \int \frac{d^4p}{(2 \pi)^4} \ \phi^{\dagger}(p)\Delta^{-1}(p) \ \phi(p).
\eeq
The non-local coordinate space expression for this action is made gauge-invariant by introducing a path-dependent Wilson line 
\beq
W_P(x,y) = P \exp \left[-i g T^a \int_x^y A_{\mu}^a(w) \ dw^{\mu}\right].
\eeq
Feynman rules are then derived by taking functional derivatives of the action with respect to the appropriate fields; this method has been used, for example, in the case of the non-local chiral-quark model \cite{NCQM}.  One must address, however, the ambiguity of defining the derivative of the line integral.  In what follows we'll see that the choice which facilitates deriving minimally coupled vertices precludes the freedom to choose any arbitrary path.

\section{Reproducing Minimal Coupling}\label{sec:Reproducing Minimal Coupling}
\setcounter{equation}{0}
In order to make progress in deriving Feynman rules from this gauge-invariant action, we now specify the appropriate definition of the derivative of the Wilson line.  We argue here that the natural choice is indeed that given by Mandelstam \cite{Mandelstam}  inasmuch as one wishes to reproduce minimal coupling.  The definition advocated here is thus such that 
\beq\label{derivative}
\frac{\partial}{\partial y^{\mu}} W_P(x,y)=-igT^a  A_{\mu}^a(y) W_P(x,y).
\eeq
It was this definition that allowed Mandelstam to reproduce the usual potential formulation of QED.  What is concluded  from the calculation in \cite{Licht1}, however, is that this expression will not hold universally; i.e. it is not satisfied by all possible paths.  Below, however, we show that implementing this definition of the derivative in the line integral approach, one derives identical Feynman rules to those derived starting from the minimal coupling prescription not only for particles, but for unparticles as well.  Alternative definitions of the derivative will generally produce more complicated Feynman rules bearing little resemblance to the familiar ones.

\subsection{Minimal Coupling for $d=1$}
We can quickly demonstrate an example of the equivalence between the minimal coupling scheme and Mandelstam's approach by considering the case $d=1$, where we expect the known result for the vertex function $\Gamma^\mu$:
\beq\label{VCan}
ig\Gamma^\mu(p,q)=ig (2 p^\mu+q^\mu).
\eeq
This method can then be compared to other approaches.  For the sake of brevity we will simply quote known results for vertex functions involving an incoming scalar with scaling dimension $d$ and momentum $p$ and a gauge field of momentum $q$.  From \cite{John}, where the relationship (\ref{derivative}) is assumed, one has for the Abelian case (suspending normalization and masses for the scalars)
\beq\label{VJ}
i g \Gamma^{\mu}(p,q,d) = -i g \frac{2 p^{\mu}+q^{\mu}}{2 p \cdot q+q^2}\left[(-(p+q)^2)^{2-d}-(-p^2)^{2-d}\right],
\eeq
which clearly reduces to Eq.~(\ref{VCan}) when $d=1$.  (In the next section we show that the minimally coupled vertex is obtained in this manner for other integer values as well).  One can also perform other non-trivial checks, e.g. that the Ward-Takahashi \cite{WT} identity is satisfied.  

From \cite{Licht1}, where a straight line path is used and relation (\ref{derivative}) is thus relinquished, one has for the same case (with $p'\equiv p+q$)
\beq\label{VL}
i g \Gamma^{\mu}(p,q,d)= -2g \left[\frac{q^{\mu}}{q^2}(p'^{2 \nu}-p^{2 \nu})+\frac{2 \nu A^{\nu -1}}{q^2}(p^{\mu} (p'\cdot q)-p'^{\mu}(p \cdot q))C_{\nu-1} \right];
\eeq
with notation specified by 
\beq
\nu\equiv 2-d;\quad A\equiv p^2-\frac{(p\cdot q)^2}{q^2}; \quad B\equiv \frac{p\cdot q}{q^2}; \nonumber 
\eeq
\beq
C_{\nu}\equiv (1+B) \cdot \, _2F_1 \left(\frac{1}{2},-\nu;\frac{3}{2};-\frac{q^2}{A}(1+B)^2\right)-B \cdot \, _2F_1 \left(\frac{1}{2},-\nu;\frac{3}{2};-\frac{q^2}{A}B^2\right).
\eeq
With this, one again recovers --- up to an overall normalization --- the usual minimal vertex function (\ref{VCan}) for particles, i.e. when $d=1$.  For $d \neq 1$, however, this turns out not to be the case: one can indeed show that for other integer values of $d$ the minimal coupling prescription is lost when starting from Eq.~(\ref{VL}).  Here then we have an explicit example of a path that doesn't satisfy Eq.~(\ref{derivative}) and thus leads to non-minimal vertices.  

While these approaches are both seen to reproduce the same result when $d=1$, we are clearly interested in all other cases, i.e. when we are actually dealing with unparticles.  In these more general cases we claim that the vertex (\ref{VJ}) derived supposing the use of a Wilson line along with the definition for its derivative given by Eq.~(\ref{derivative}) is the natural extension to unparticles of the usual minimal coupling prescription.

\subsection{Generalized Minimal Coupling}
We would now like to prove that the vertex function for scalar unparticles given in \cite{John} matches that derived from a simple application of the minimal coupling prescription for other integer values of $d$.  

We take $n$ to denote the power of the Lorentz-invariant derivative term, and thus consider the action
\beq\label{action}
S_n = -\int d^4 x \ \phi^{\dagger}(x)(D_\mu D^\mu)^n \phi(x)
\eeq
where
\beq
D_\mu \equiv \partial_\mu + igA_\mu
\eeq
defines minimal coupling, and the gauge field is understood to be analyzed at the same point as the scalar fields.

The vertex of interest is defined by
\beq\label{vertex}
ig\Gamma^{\mu}_n(p,q) &\equiv& \left.\frac{i\delta^3 S_n}{\delta A^{\mu}(q) \delta \phi^\dagger(p+q)\delta \phi(p)}\right\vert_{A^\mu = 0}
\eeq 
and we wish to show explicitly that if $n$ is an integer greater than or equal to zero, then this definition along with Eq.~(\ref{action}) produces the expression
\beq\label{vertassert}
ig\Gamma^{\mu}_n(p,q) &=& -ig \frac{2p^\mu + q^\mu}{2p \cdot q + q^2} \left[(-(p+q)^2)^n - (-p^2)^n \right] \nonumber \\
& =& -ig (-1)^n \frac{2p^\mu + q^\mu}{2p \cdot q + q^2} \left[(p+q)^{2n} - p^{2n} \right] ,
\eeq
i.e. that it exactly matches Eq.~(\ref{VJ}).

\bigskip

\noindent To begin the inductive proof, we first show that the assertion is true for $n = 1$. In this case, the right hand side of (\ref{vertassert}) reduces to 
\beq\label{eqn:n=1}
ig\Gamma^{\mu}_1(p,q) = ig(2p^\mu + q^\mu).
\eeq
We also have
\beq
\phi^\dagger(x)(D^2)\phi(x) &=& \phi^\dagger(x)(\partial_\mu + igA_\mu)(\partial^\mu + igA^\mu)\phi(x) \nonumber \\
& \rightarrow &\phi^\dagger \left(-p^2 + ig(2ip^\mu + iq^\mu)A_\mu -g^2 A_\mu A^\mu \right) \phi,
\eeq
so that from Eq.~(\ref{vertex})
\beq
ig\Gamma^{\mu}_1(p,q)  = ig(2p^\mu + q^\mu), 
\eeq
in agreement with Eq.~(\ref{eqn:n=1}).

\bigskip

\noindent Now we analyze the case $n \rightarrow n+1$.  The right hand side of Eq.~(\ref{vertassert}) is given in this case by
\beq\label{n+1}
ig\Gamma^{\mu}_{n+1}(p,q) = -ig (-1)^{n+1} \frac{2p^\mu + q^\mu}{2p \cdot q + q^2}\left[(p+q)^{2(n+1)} - p^{2(n+1)} \right],
\eeq
and now we have
\beq
D^{2(n+1)}\phi = D^{2n}D^2\phi = D^{2n}\left[\partial^2 + ig(A^\mu \partial_\mu + \partial_\mu A^\mu) - g^2 A_\mu A^\mu \right] \phi.
\eeq
Going to momentum space:
\beq
D^{2(n+1)}\phi = -p^2D^{2n}\phi - g(2p^\mu + q^\mu)D^{2n}(A_\mu \phi) + \mathcal{O}(A^2)
\eeq
where terms $\mathcal{O}(A^2)$  will drop out upon differentiating the action.

We also know that
\beq
D^{2n}(A_\mu \phi) = \partial^{2n}(A_\mu \phi) + \mathcal{O}(A^2)
\eeq
Again going to momentum space, and using the fact that $\partial^2 (A_\mu \phi) = -(p^2+2p \cdot q + q^2)(A_\mu \phi) = -(p+q)^2 (A_\mu \phi)$, we find that
\beq
\label{eqn:parAphi}
\partial^{2n}(A_\mu \phi) = (-1)^n (p+q)^{2n}(A_\mu \phi)
\eeq
Using this result, we have
\beq
\phi^\dagger D^{2(n+1)}\phi = -p^2 \phi^\dagger D^{2n}\phi - g(-1)^n(2p^\mu + q^\mu)(p+q)^{2n} \phi^\dagger A_\mu \phi + \mathcal{O}(A^2)
\eeq
Therefore using Eq.~(\ref{vertassert}), we now find
\beq
\label{eqn:n+1}
\left. \frac{i\delta^3 S_{n+1}}{\delta A^{\mu}(q) \delta \phi^\dagger(p+q)\delta \phi(p)}\right\vert_{A^\mu = 0} &=&\left. -p^2\frac{i\delta^3 S_{n}}{\delta A^{\mu}(q) \delta \phi^\dagger(p+q)\delta \phi(p)}\right\vert_{A^\mu = 0} \nonumber \\
&& \qquad \qquad \qquad+ ig(-1)^n(2p^\mu + q^\mu)(p+q)^{2n} \nonumber \\
& =& igp^2(-1)^n \frac{2p^\mu + q^\mu}{2p \cdot q + q^2} \left[(p+q)^{2n} - p^{2n} \right] \nonumber \\
&& \qquad \qquad \qquad+ ig(-1)^n(2p^\mu + q^\mu)(p+q)^{2n}.
\eeq
Finally, combining these two terms we have
\beq
\left.\frac{i\delta^3 S_{n+1}}{\delta A^{\mu}(q) \delta \phi^\dagger(p+q)\delta \phi(p)}\right\vert_{A^\mu=0} = -ig(-1)^{n+1} \frac{2p^\mu + q^\mu}{2p \cdot q + q^2}\left[(p+q)^{2(n+1)} - p^{2(n+1)} \right].
\eeq
This matches the result in Eq.~(\ref{n+1}) and thus completes the proof of our original assertion.

\section{Conclusions}\label{sec:Conclusions}
\setcounter{equation}{0}

Minimal coupling has proven phenomenologically successful in describing elementary particles. In the absence of compelling theoretical or phenomenological reasons to do otherwise, it seems most natural to generalize the minimal coupling prescription to unparticles. We have shown that the vertex produced by Terning et al \cite{John} does this by allowing non-integer $d$, and reduces to minimal coupling for integer $d$. Although the definition for the derivative supported here is a specific restriction on the integral operator we use to ensure gauge invariance, it is the one that produces the expected vertices and is a useful construction in this regard.  While it may be possible to find an alternative integral operator which satisfies this crucial derivative criterion for all paths, we will not pursue it here.

\section*{Acknowledgments}

We thank John Terning and A.~Lewis Licht  for useful discussions
and comments.  This work is 
supported by the US Department of Energy under contract DE-FG03-91ER40674.

Note added:  While this work was in preperation, another alternative construction \cite{Licht2} -- not relying on the use of a Wilson line -- appeared which reproduces the vertex advocated here.

\end{document}